\documentclass[10pt]{article}

\usepackage{graphics}
\usepackage{graphicx}

\usepackage[a4paper, left=35mm,right=35mm,top=34mm,bottom=34mm]{geometry}
\usepackage[utf8]{inputenc}
\usepackage[T1]{fontenc}
\usepackage[english]{babel}

\usepackage{enumerate}
\usepackage{graphicx}
\usepackage{hyperref}
\hypersetup{
    colorlinks=true,
    linkcolor=blue,
    filecolor=magenta,      
    urlcolor=cyan,
}

\usepackage{mathtools,amsthm,amssymb,amsfonts}
\usepackage{algorithm}
\usepackage{algorithmic}
\makeatother
\theoremstyle{plain}

\theoremstyle{definition}

\theoremstyle{remark}

\usepackage{caption} 
\usepackage{fancyhdr}

\newcommand{\abs}[1]{\left\lvert#1\right\lvert}
\newcommand{\norm}[1]{\left\lVert#1\right\rVert}

\author{
  {\normalsize Qinmeng Zou}\thanks{CentraleSup\'elec, Universit\'e Paris-Saclay, France.}
  \and
  {\normalsize Fr\'ed\'eric Magoul\`es}\thanks{CentraleSup\'elec, Universit\'e Paris-Saclay, France
    (correspondence, frederic.magoules@hotmail.com).}
		}
\title{Convergence Detection of Asynchronous Iterations based on Modified Recursive Doubling}
\date{}

\begin{document}
\maketitle
\thispagestyle{fancy}

\begin{abstract}
\noindent This paper addresses the distributed convergence detection problem in asynchronous iterations.
A modified recursive doubling algorithm is investigated in order to adapt to the non-power-of-two case.
Some convergence detection algorithms are illustrated based on the reduction operation.
Finally, a concluding discussion about the implementation and the applicability is presented.
\end{abstract}


\section{Introduction}

Consider the following linear system
\[
Ax = b,
\]
where $A\in\mathbb{R}^{n\times n}$ and $b\in\mathbb{R}^n$.
A splitting
\[
A = M - N,
\]
yields an iterative scheme
\[
x^{k+1} = Tx^k + c,
\]
where $k\in\mathbb{N}$, $T = M^{-1}N$ and $c = M^{-1}b$.
Generally, this scheme is well suited for parallel computing of the form
\[
x_i^{k+1} = T_ix^k + c_i,\quad i\in\{1,\ \dots,\ p\},
\]
where $p$ is the number of processors.
Here, $x_i$ and $c_i$ might be two values or two smaller vectors.
Similarly, $T_i$ might be a row vector or a smaller matrix.
They are distributed in different processors.
However, a specific point is required at the end of each iteration to synchronize between the processors.
The waste of time may be significant in the case of unbalanced working load and node failure, which gives rise to the asynchronous iterative methods.
The asynchronous iterative scheme has been proposed by Chazan and Miranker \cite{Chazan1969} for the solution of linear equations and generalized by several researchers (see, e.g., \cite{Miellou1975, Baudet1978, ElTarazi1982, Bertsekas1989a}) for the general problem
\[
x^{k+1} = f\left(x^k\right),
\]
where $f$ is a fixed point mapping.
The asynchronous iterative scheme is shown as follows
\[
x_i^{k+1} =
\begin{cases}
f_i\left(x_1^{\tau_{i,1,k}},\ \dots,\ x_p^{\tau_{i,p,k}}\right), & i\in P^{k}, \\
x_i^{k}, & i\notin P^{k},
\end{cases}
\]
where $\tau_{i,j,k}\le k$ is a sequence of iterations with retards for each element $j$ in each processor $i$, and $P^{k}\subset\{1,\ \dots,\ p\}$ is a sequence of subsets of processor numbers.
In this case, processors are not required to wait for receiving all messages and allowed to keep on their own pace.
We often add the following conditions to better investigate the chaotic process
\[
\begin{cases}
\abs{\left\{k\in\mathbb{N} \mid i\in P^{k}\right\}} = +\infty, & \forall i \in \{1, \dots, p\}, \\
\underset{k\to+\infty}{\lim}\tau_{i,j,k} = +\infty, & \forall i,j \in \{1, \dots, p\},
\end{cases}
\]
where $\abs{.}$ is the cardinality of a set that measures the number of elements.
It means that no processors should be abandoned forever and more and more recent values should be used.

Asynchronous iterative algorithms must terminate after a finite number of iterations, as suggested in \cite{Savari1996}.
Thus, a practical implementation involves a set of admissible solutions $S$, such that
\[
x^*\in S,
\]
where $x^*$ is a solution vector.
We would like to find a vector $\bar{x}$ established by the components from each processor, and we have to evaluate $\bar{x}\in S$. If true, then $x^* = \bar{x}$; otherwise, continue the computation, as well as the evaluation.
Thus, the termination condition can be expressed by a residual evaluation
\[
\norm{f\left(\bar{x}\right) - \bar{x}} < \epsilon,\quad \epsilon>0,
\]
where $\norm{.}$ is a norm, $\epsilon$ is a well-chosen threshold.
$\bar{x}$ is given as an arbitrary combination of local components
\[
\bar{x} = \left(x_1^{k_1},\ \dots,\ x_p^{k_p}\right),\quad k_1,\ \dots,\ k_p\in\mathbb{N}.
\]
The major problem of termination detection is how to collect $x_i^{k_i}$ and execute the evaluations.

Recently, several developments for the asynchronous iterations have been proposed in different domains, such as domain decomposition methods \cite{Magoules2017a, Magoules2018c, Magoules2018e}, convergence detection methods \cite{Miellou2008, Magoules2018a}, and programming libraries \cite{Magoules2017b, Magoules2018b}.
In this paper, we propose a modified recursive doubling algorithm applied to the non-blocking collective communication, which is addressed in the next section.
In Section~\ref{sec:3}, we present some convergence detection strategies based on our new method.
Finally, further discussion is given in Section~\ref{sec:4} about the implementation and performance.

\section{Modified Recursive Doubling}
\label{sec:2}

Recall that $p$ is the number of processors.
We define a $\mu_0\in\mathbb{N}$ such that
\[
p_0 = 2^{\mu_0} \le p < 2^{\mu_0+1},
\]
where $p_0$ denotes a pivot.
Note that for the parallel iterative methods, the \texttt{Allreduce} function is very useful because we need to collect residual values from different processors.
For the asynchronous iterations, however, the collective operations should be not only efficient, but also performed in a non-blocking way.

\begin{figure}[!ht]
\centering
\includegraphics[width=4.1in]{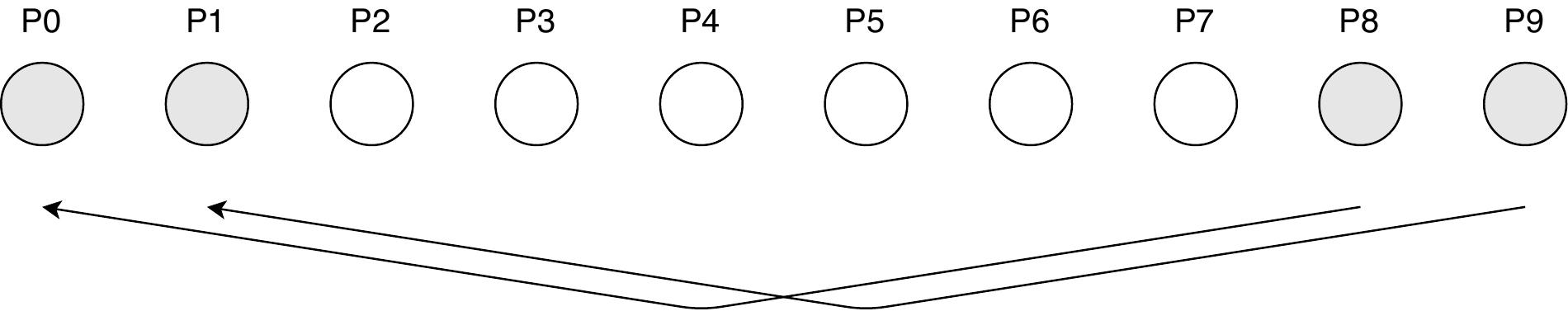}
\caption{Backward shift}
\label{fig:1}
\end{figure}

\begin{figure}[!ht]
\centering
\includegraphics[width=4.1in]{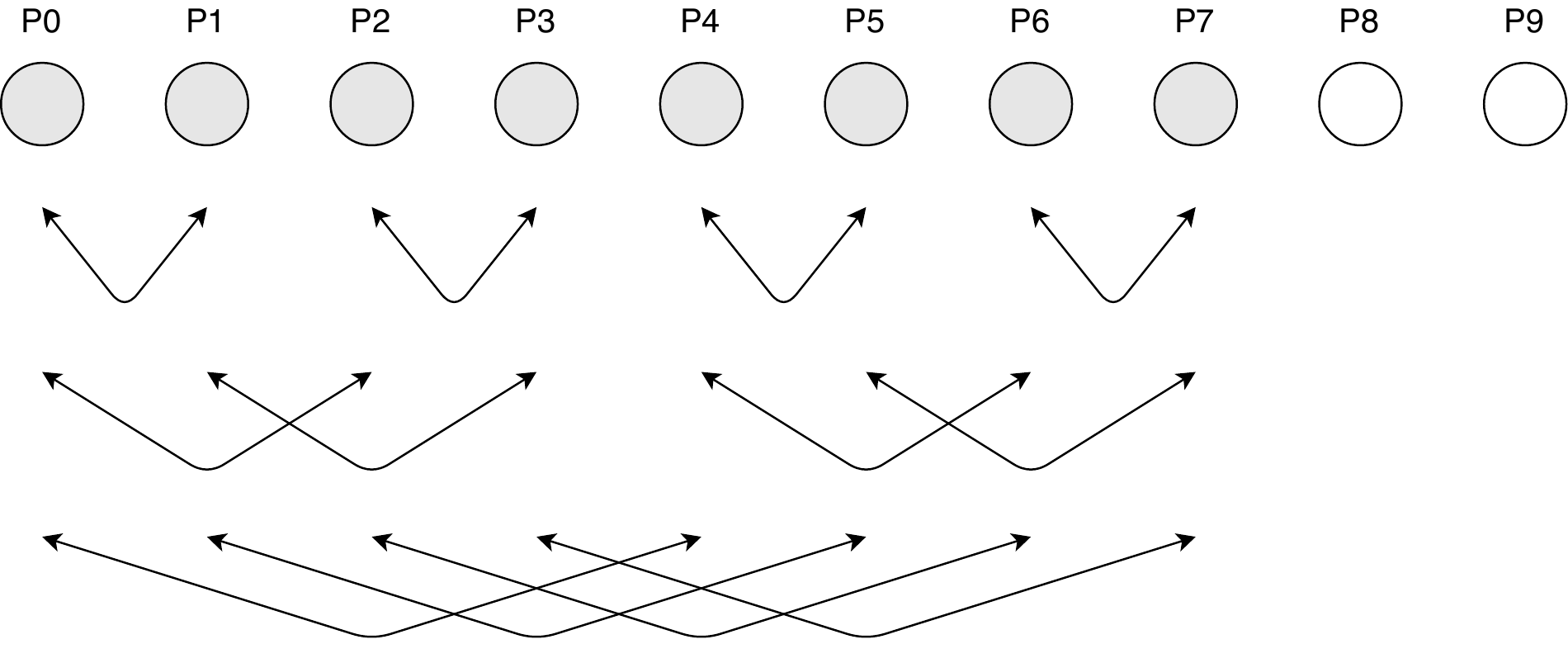}
\caption{Recursive doubling}
\label{fig:2}
\end{figure}

\begin{figure}[!ht]
\centering
\includegraphics[width=4.1in]{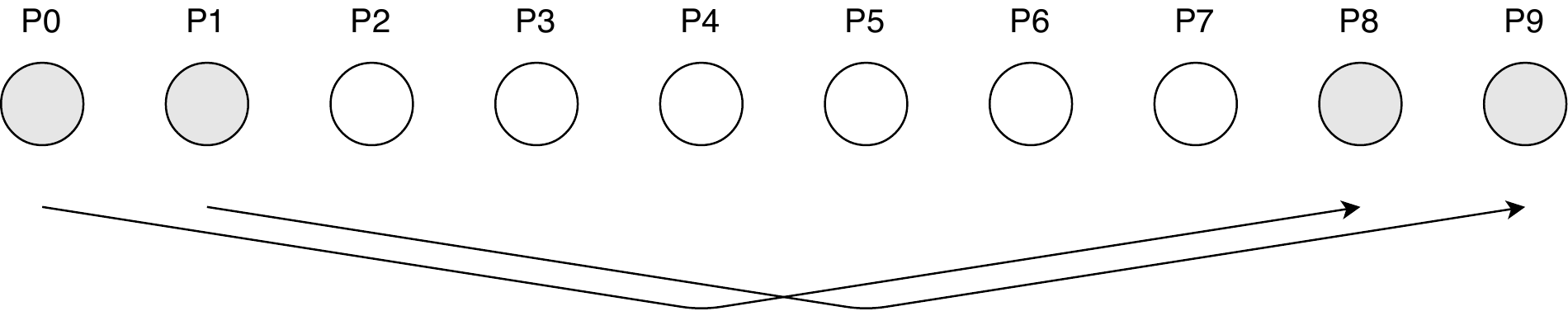}
\caption{Forward shift}
\label{fig:3}
\end{figure}

Traditional recursive doubling (see, e.g., \cite{Thakur2005}) is one of the possible algorithms for the \texttt{Allreduce} function.
It involves a power-of-two number of processors, which adapts only to some special situations.
We consider a modified version of recursive doubling, in which a backward shift and a forward shift are required in the general case.
The first step is sending data from the extra processors to the first several processors, called backward shift, illustrated in Figure~\ref{fig:1}.
During this process, we proceed the corresponding arithmetical operations, such as summation, maximization, and minimization.
Then, the recursive doubling algorithm is proceeded only within the power-of-two processors to exchange data and execute reduction operation as shown in Figure~\ref{fig:2}.
Finally, a forward shift is proceeded to send back the final data to the extra processors as illustrated in Figure~\ref{fig:3}, which is indeed the inverse operation of the first shift.

\begin{figure}[!ht]
\centering
\includegraphics[width=4.6in]{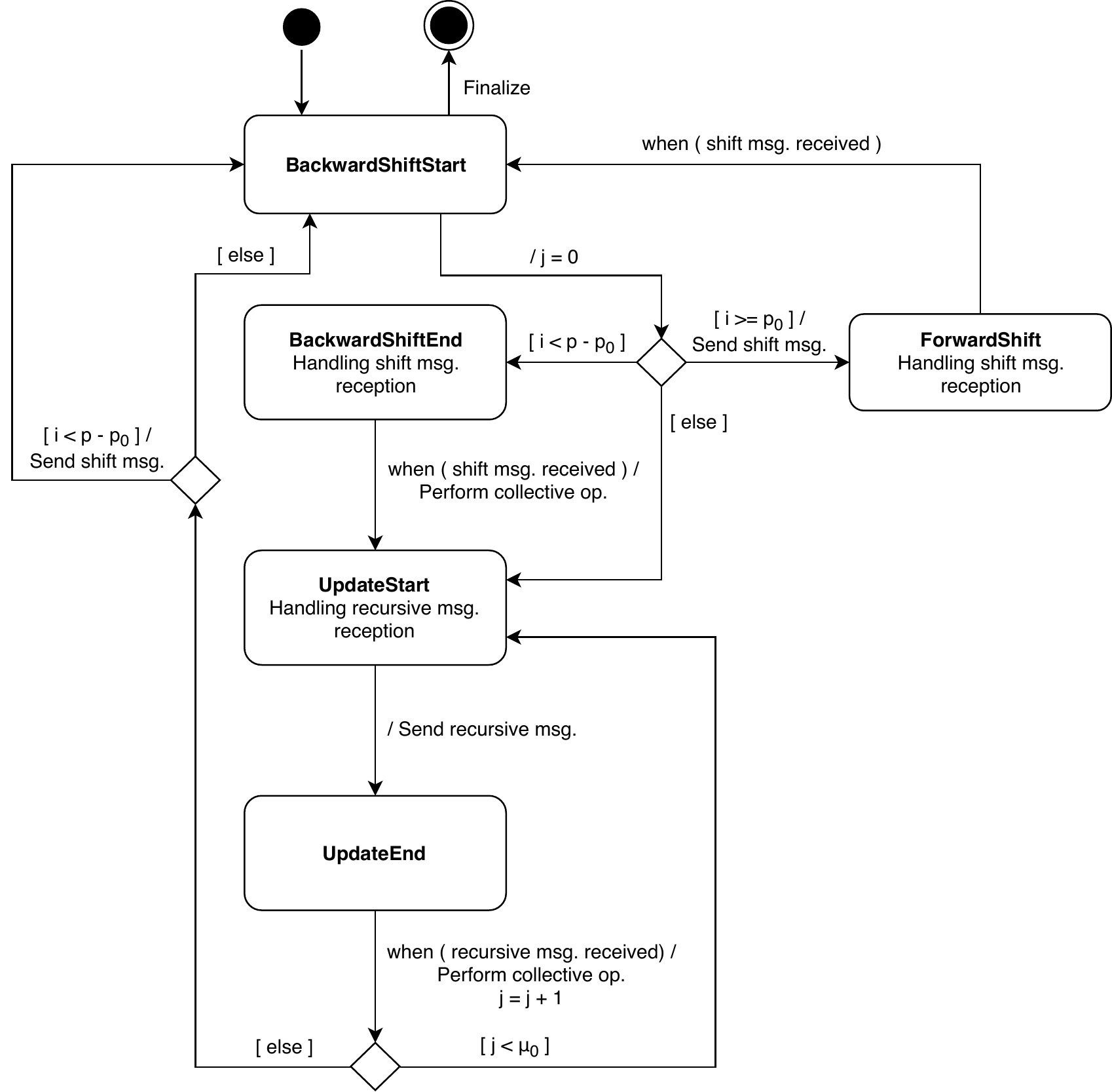}
\caption{Statechart of a non-blocking \texttt{Allreduce} function}
\label{fig:4}
\end{figure}

Asynchronous iterations require non-blocking communication, which can be implemented is several ways.
For example, we might prefer to create a new thread for a desired collective function, and then design its behavior by some external interface functions;
we could also create a state-based interface that should be invoked repeatedly in user applications, in which some lightweight functions act as different states in the life cycle of a collective operation.
Here, we adopt the latter and give an example of state diagram depicted in Figure~\ref{fig:4}, which implements a non-blocking \texttt{Allreduce} function.
From the picture, we can see that each cycle begins with the backward shift operation.
If the rank of processor belongs to the extra range, it sends data and enters into the forward shift state.
In other cases, the relative processors must enter into the updating loop, which involves all the processors having a rank smaller than the pivot $p_0$.
Finally, a forward shift process is executed to gather the final results to the non-power-of-two processors.
Notice that the first several processors within the exponential area engage as well in the shift subroutines.

The amount of data exchanged by each processor depends on the way of collecting residual values.
We need exactly $\log p_0 + 2$ steps to finish a cycle in the synchronous case.
If there is only a floating point residual value being exchanged in each processor, then totally $p_0\log p_0 + 2(p-p_0)$ data are exchanged in each cycle.
In asynchronous mode, this number keeps the same.
However, processors wait no longer the others and conduct iterations on their own pace.

\section{Convergence Detection Algorithms}
\label{sec:3}

We could also develop other collective operations based on the backward-forward recursive doubling algorithm.
In practice, however, these functions are rarely used in the context of asynchronous iterations because we expect to exploit the most recent values as much as possible, which favors the point-to-point operations like \texttt{Send} and \texttt{Recv} functions.
On the other hand, the residual collection requires intrinsically an \texttt{Allreduce} operation.
Therefore, we address the convergence detection problem in terms of the non-blocking reduction.

We consider first an inexact residual collection strategy that involves only the \texttt{Allreduce} function, depicted as follows.
\begin{algorithmic}
\STATE $\texttt{res\_norm} = \texttt{res\_thresh}$
\STATE $\texttt{res\_loc} = \texttt{res\_thresh}$
\STATE $\texttt{res\_glb} = \texttt{res\_thresh}$
\STATE $\texttt{flag} = 1$
\WHILE{$\texttt{res\_norm} \ge \texttt{res\_thresh}$}
\STATE $z_i = x_i$
\STATE $\texttt{Compute}(x_i,\ A_i,\ b_i,\ x)$
\STATE $\texttt{Send}(x_i)$
\STATE $\texttt{Allreduce}(\texttt{res\_glb},\ \texttt{res\_loc},\ \texttt{flag})$
\IF{\texttt{flag}}
\STATE $\texttt{res\_norm} = \texttt{res\_glb}$
\STATE $\texttt{res\_loc} = \norm{x_i-z_i}_\infty$
\STATE $\texttt{flag} = 0$
\ENDIF
\STATE $\texttt{Recv}(x)$
\ENDWHILE
\end{algorithmic}
We mention here that although such algorithm is not exact, it might be efficient due to the simplicity and still has an acceptable precision.
In the algorithm, we take the maximum norm as an example to compute residual and omit some function parameters, e.g., the arithmetic operation of reduction.
Unlike the message-passing standard \cite{MPIF1994}, our implementation is based on the state that requires the function invocation repeatedly, not just a request handler.
The \texttt{Compute} function could be any appropriate iterative algorithm, such as Jacobi method or gradient method (see, e.g., \cite{Bertsekas1989a}).
This is inexact because \texttt{res\_loc} might not be monotone all over the iterations.
Sometimes global residual indicates a convergence signal but local residual rises instead due to the retard term.

Now we give a second version that leads to an exact solution in view of the residual collection, shown as follows.
\begin{algorithmic}
\STATE $\texttt{res\_norm} = \texttt{res\_thresh}$
\STATE $\texttt{sflag} = 1$
\STATE $\texttt{eflag} = 0$
\WHILE{$\texttt{res\_norm} \ge \texttt{res\_thresh}$}
\STATE $\texttt{Compute}(x_i,\ A_i,\ b_i,\ x)$
\STATE $\texttt{Send}(x_i)$
\IF{\texttt{sflag}}
\STATE $\texttt{Snapshot}(\bar{x},\ x_i,\ \texttt{eflag})$
\IF{\texttt{eflag}}
\STATE $z_i = \bar{x}_i$
\STATE $\texttt{Compute}(\bar{x}_i,\ A_i,\ b_i,\ \bar{x})$
\STATE $\texttt{res\_loc} = \norm{\bar{x}_i-z_i}_\infty$
\STATE $\texttt{sflag} = 0$
\STATE $\texttt{eflag} = 0$
\ENDIF
\ELSE
\STATE $\texttt{Allreduce}(\texttt{res\_glb},\ \texttt{res\_loc},\ \texttt{eflag})$
\IF{\texttt{eflag}}
\STATE $\texttt{res\_norm} = \texttt{res\_glb}$
\STATE $\texttt{sflag} = 1$
\STATE $\texttt{eflag} = 0$
\ENDIF
\ENDIF
\STATE $\texttt{Recv}(x)$
\ENDWHILE
\end{algorithmic}
This algorithm involves a distributed snapshot process that generates a consistent solution buffer \cite{Chandy1985} (see also, e.g., \cite{Savari1996, Magoules2018a}).
The snapshot algorithm first sends $x_i$ to the processors that depend on $x_i$.
In this situation, we call them dependent neighbors; then, processor $i$ begins to wait for the necessary data from some other processors, which are called essential neighbors \cite{Bertsekas1989a}; finally, it returns a collection of essential data that are used for the residual computation.
Here we simplify the process by assuming that the communication follows an ``all-to-all'' pattern, which implies that both dependent neighbors and essential neighbors are all the other processors so that they are the same.
For the general case, the algorithm would be similar.
We first set $\texttt{sflag}=1$ to enable snapshot process.
Then, we compute \texttt{res\_loc} when snapshot finishes and set all $\texttt{sflag}=0$, which enables the reduction process.
Finally, \texttt{Allreduce} is called repeatedly that is exactly the first algorithm, except that this time we keep a set of consistent data that provides an exact result.

\section{Further Discussion}
\label{sec:4}

In this section we first discuss the implementation issue of the convergence detection algorithms.
Here we take the message passing interface (MPI) standard as an example.
Notice that in order to implement a non-blocking function, we should execute the relative instructions in an independent thread, which involves an explicit construction or an implicit invocation.
We choose the latter and invokes the non-blocking point-to-point instructions to exchange messages.
We can use the external interface functions to generate a non-blocking function under the name of generalized requests.
In current version, the two main functions are \texttt{MPI\_Grequest\_start} and \texttt{MPI\_Grequest\_complete}.
In the next version, these functions will be redefined in order to provide a more flexible interface.

Notice that if the number of processors falls on the power-of-two case, the iterations in Figure~\ref{fig:4} will jump over all the shift steps appropriately.
Such case has been proven very efficient in several situations \cite{Thakur2005}, whereas our algorithm can benefit from it as well.
On the other hand, our $\texttt{Send}$ operation is implemented in a blocking mode because it causes rarely a negative impact in practice on the efficiency.
We could avoid wasting time by switching it to non-blocking mode without changing so much codes.

\begin{figure}[!t]
\centering
\includegraphics[width=4.1in]{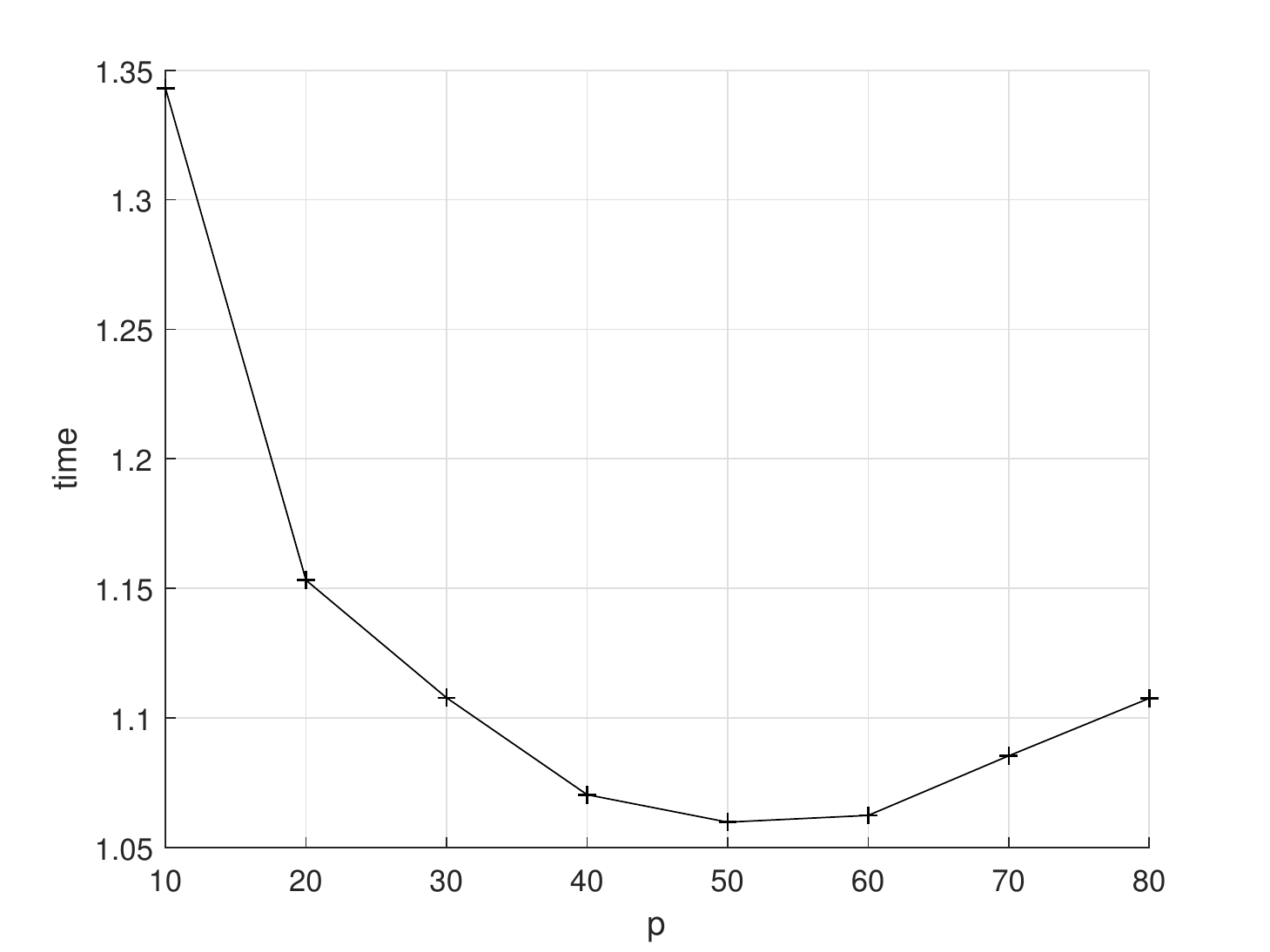}
\caption{Asynchronous iterations in a concentrated environment}
\label{fig:5}
\end{figure}

Finally, we mention here that our algorithm is suitable for a relatively ``close'' distributed environment; otherwise, there might be a great deal of communication operations exchanging data between long-distance nodes, which increases the transfer time.
In such case, a tree-based algorithm is preferred.
However, asynchronous iterations may not exhibit advantages in a completely local cluster, even perform sequentially like synchronous scheme with much more ongoing messages.
Consider a two-point boundary value problem with an asynchronous relaxation solver \cite{Chazan1969}.
We implement the mathematical operations by Alinea \cite{Magoules2015a} and the asynchronous iterations by JACK \cite{Magoules2018b}, which have been proven very efficient for the large-scale scientific computing \cite{Magoules2015c, Magoules2015d, Magoules2016b}.
The finite difference scheme is adopted for the discretization.
The matrix dimension $n=10000$ and $b$ is chosen arbitrarily from $-10$ to $10$.
Results are shown in Figure~\ref{fig:5}.
We observe that the iteration curve shows synchronous behavior that exists a bottleneck within a specific range of processors.
The experiment was performed on a cluster of Intel Xeon CPU E5-2670 v3, connected by FDR Infiniband network with 56 Gbit/s, which is concentrated and favors synchronous iterations.
Furthermore, an ``all-to-all'' algorithm generates huge amounts of messages in the asynchronous mode, which makes the network too messy to be efficient.
In this case, we prefer the traditional synchronous iterative scheme, even for large-scale parallel computing.

\section*{Acknowledgment}
This work was supported by the French national programme LEFE/INSU and the project ADOM (M\'ethodes de d\'ecomposition de domaine asynchrones) of the French National Research Agency (ANR).

\bibliography{ref}
\bibliographystyle{abbrv}

\end{document}